\documentclass[12pt, a4paper]{article}
\usepackage{jcappub}
\pdfoutput=1
\usepackage{amsmath}
\usepackage{graphicx}

\def\P{\mathcal{P}}

\newcommand{\PRE}[1]{}
\newcommand{\Expect}[1]{\left\langle #1 \right\rangle}

\newcommand{\be}{\begin{equation}}
\newcommand{\ee}{\end{equation}}
\newcommand{\bea}{\begin{eqnarray}}
\newcommand{\eea}{\end{eqnarray}}
\newcommand{\beas}{\begin{eqnarray*}}
\newcommand{\eeas}{\end{eqnarray*}}
\newcommand{\V}[1]{\vec{#1}}

\title{Local Scale-Dependent Non-Gaussian Curvature Perturbations at Cubic Order }

\author{Joseph Bramante,}
\author{Jason Kumar}

\affiliation{Department of Physics and Astronomy, University of
Hawaii, \\ 2505 Correa Rd., Honolulu HI, USA}

\emailAdd{jkumar@hawaii.edu}
\emailAdd{bramante@hawaii.edu}

\abstract{We calculate non-Gaussianities in the bispectrum and trispectrum arising from the cubic term
in the local expansion of the scalar curvature perturbation.  We compute to three-loop order
and for general momenta.
A procedure for evaluating the leading behavior of the resulting loop-integrals is developed and discussed.
Finally, we survey unique non-linear signals which could arise from the cubic term in the squeezed limit.
In particular, it is shown that loop corrections can cause $f_{NL}^{sq.}$ to change sign as the momentum
scale is varied. There also exists a momentum limit where $\tau_{NL} <0$ can be realized.}
\arxivnumber{1107.5362}

\begin{document}
\maketitle

\section{Introduction}

A variety of new experiments are poised to probe the non-Gaussianity of primordial
density fluctuations with unprecedented accuracy.  These experiments include the Planck
satellite~\cite{Planck} alongside a variety of probes of large scale structure formation~\cite{LSS}.
As a result non-Gaussianity can be probed at a wide variety of scales.  It thus becomes important to
determine which types of models can yield non-Gaussian fluctuations which can be probed
at these experiments, and conversely, how measurements from these experiments can be used
to constrain inflationary models.

Along this line, considerable work has focused on the local ansatz~\cite{localansatz} for non-Gaussian
fluctuations, which is realized in many inflationary models.  In particular it was shown
that non-Gaussianity of the trispectrum is related to non-Gaussianity of the bispectrum
through the inequality
\bea
\tau_{NL} \geq \left({6 \over 5} f_{NL} \right)^2
\eea
at tree-level~\cite{Suyama:2007bg}, provided the curvature scalar is expanded to quadratic order in Gaussian fields.
This became an important constraint, because it implied that measurements at upcoming
experiments could potentially rule out the applicability of the local ansatz.

But subsequently it was shown in~\cite{Sugiyama:2011jt} that
this relation is modified to
\bea
\tau_{NL} \geq {1\over 2} \left({6 \over 5} f_{NL} \right)^2
\eea
if one expands to quartic order and includes one-loop corrections.  This raises
the important question of whether there is indeed a rigid constraint on the trispectrum
in the local ansatz, or whether any such constraint can be violated if one computes
to sufficiently high loop order.  This is especially interesting because it has been
shown that in reasonable models loop contributions can dominate over tree-level
contributions~\cite{Kumar:2009ge}.

In this work, we consider a local ansatz for non-Gaussianity in a multi-field
model of inflation (as non-Gaussianity is expected to be very small in single-field
models with a standard kinetic term~\cite{Maldacena:2002vr}), expanded to cubic order.
We calculate the power spectrum, bispectrum and trispectrum up to three-loop order
(no higher loop diagram exists for $n$-point functions with $n \leq 4$ at cubic order
in the expansion).  We will find a variety of interesting features which arise from
the non-trivial scale-dependence of loop corrections (see also~\cite{scaledep}).
In particular, we find that $f_{NL}$ can change sign with momentum scale in the
squeezed limit.  Moreover, $\tau_{NL}$ can be negative in a particular limit of
the external momenta.

In section 2, we review the local ansatz and the various local momentum shapes which
are generated at tree-level.  In section 3, we describe the formalism for calculating
loop diagrams.  In section 4, we present results for the computation of the power
spectrum, bispectrum and trispectrum to cubic order (detailed calculations are presented
in the appendices).  We conclude with a discussion of our results in section 5.

\section{Local Ansatz For Curvature Perturbations}

The local ansatz for the curvature scalar $\zeta$ amounts to the assumption that $\zeta (t, {\bf x})$ can
be written as a non-linear function of one or more Gaussian scalar fields $\phi_i$, all evaluated
at the same space-time point $(t, {\bf x})$.  For simplicity, we will assume that there are only
two fundamental scalars of interest, the inflaton $\phi$ and an additional field $\chi$.

It was shown in~\cite{Maldacena:2002vr} that single-field models of inflation with a standard kinetic
term will only yield very small non-Gaussianities.  The argument is intuitively quite simple: non-Gaussianities
are typically generated by some type of non-linearity in the interaction potential.  But the inflaton
is constrained to have an extremely flat potential, so the types of non-linearities which could easily
generate non-Gaussian curvature fluctuations are constrained by the slow-roll conditions, and non-Gaussianities
in single-field models of inflation are thus proportional to the slow-roll parameters.

Of course, there are several ways to avoid this argument, and the one we will focus on is multi-field
inflation~\cite{multifield}.  In this case, it is assumed that the inflaton's interactions are largely Gaussian with
any non-linearities suppressed by slow-roll parameters.  However, there are additional scalar fields which
can have significant non-linear interactions, since their interactions are not constrained by the
slow-roll conditions.  It is these interactions which then feed into scalar curvature perturbations,
providing observable non-Gaussianity.

There are many inflationary models which generate non-Gaussianity which is approximately local (see,
for example,~\cite{multifield,curvaton,inhomogeneous} ).  We will
not focus on any particular model, however, instead assuming the phenomenological ansatz
\bea
\zeta (t, {\bf x}) &=& C_1 \phi (t, {\bf x})
+ A_1 \chi (t, {\bf x}) + {1\over 2} A_2 \left(\chi (t, {\bf x})^2 - \langle \chi^2 \rangle \right)
+ {1\over 6} A_3 \chi (t, {\bf x})^3 +\ldots
\\
\zeta_{\vec{k}} &=& C_1 \phi_{\vec{k}}
+ A_1 \chi_{\vec{k}} + {1\over 2} \int{d^3 k' \over (2\pi)^3}
\chi_{\vec{k}'} \chi_{\vec{k}-\vec{k}'}
+{1\over 6} A_3 \int{d^3 k' \over (2\pi)^3}{d^3 k'' \over (2\pi)^3}
\chi_{\vec{k}'} \chi_{\vec{k}''}
\chi_{\vec{k}-\vec{k}'-\vec{k}''}
\nonumber\\
&\,& +\ldots
\eea
In general, of course, the coefficients $C_1$, $A_i$ can depend weakly on $k$, though this scale-dependence
is constrained by bounds on the running of the power spectrum.  We will assume that these coefficients are
scale-independent.  We also assume that the lower bound on momentum is given by $L^{-1}$, where $L$ is the
size of the observable universe.  Any momenta smaller than $L^{-1}$ correspond to wavelengths larger than the
size of the observable universe, which cannot be distinguished from a constant zero mode.  A change in the
scale of this cutoff $L$ will change the value of the coefficients $A_i$, but not the value of the complete
$n$-point correlator.  Note also that for multi-field models, the scalar curvature fluctuation can exhibit
super-Hubble evolution.  We will assume, however, that these effects are small.

By assumption both $\phi$ and $\chi$ are Gaussian fields, and their 2-point correlators are
given by
\bea
\langle \phi_{\vec{k}} \phi_{\vec{k}'} \rangle =
\langle \chi_{\vec{k}} \chi_{\vec{k}'} \rangle &=&
(2\pi)^3 \delta^3 (\vec{k}+\vec{k}')
{2\pi^2 {\cal P} \over k^3}
\nonumber\\
&=&
(2\pi)^3 \delta^3 (\vec{k}+\vec{k}')
2\pi^2 {\cal P} F(\vec{k}, \vec{k}')
\eea
where
\bea
{\cal P} &=& \left( {H \over 2\pi} \right)^2
\nonumber\\
F(\vec{k}, \vec{k}') &=& {1\over k^3}
\eea

\begin{figure}[h]
\centering
\includegraphics[width=2.5in]{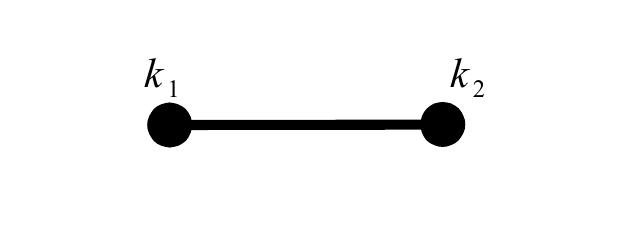}
\caption{Tree-level representation of the power spectrum.}
\label{powerspectree}
\end{figure}

\begin{figure}[h]
\centering
\includegraphics[width=4in]{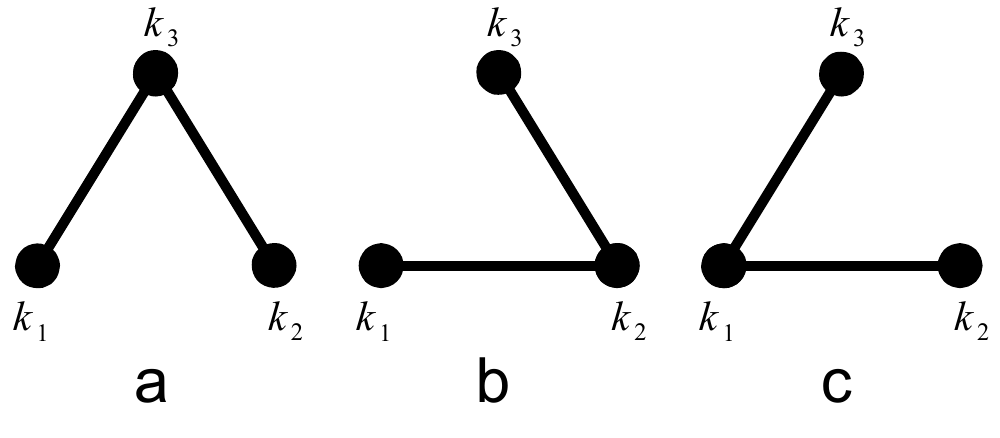}
\caption{Tree-level contributions to the bispectrum.}
\label{FNLtree}
\end{figure}

We see that correlators of the curvature scalar can be written entirely in
terms of the above Gaussian correlators.
In particular, the lowest order contribution to the 2-point correlator of the
curvature scalar is given by
\bea
\langle \zeta_{\vec{k}} \zeta_{\vec{k}'} \rangle &=&
(2\pi)^3 \delta^3 (\vec{k}+\vec{k}')
{2\pi^2 {\cal P} (C_1^2 +A_1^2) \over k^3}
\eea
It is easiest to visualize this with a diagram (see also \cite{Yokoyama:2008by}),
in which each vertex corresponds
to an insertion of $\zeta_{\vec{k}}$, while each line corresponds to a 2-point correlator
of Gaussian fields.  The two-point correlator is then represented by Fig.~\ref{powerspectree}.
Similarly, the lowest order diagrams which contribute to the bispectrum are
given in Fig.~\ref{FNLtree}.  Note that only Gaussian correlators of $\chi$ contribute
to these diagrams.
The lowest-order contribution to the 3-point correlator is given by
\bea
\langle \zeta_{\vec{k}_1} \zeta_{\vec{k}_2} \zeta_{\vec{k}_3} \rangle
&= & A_1^2A_2(2\pi^2\mathcal{P})^2(2\pi)^3
\delta^3(\Sigma_i\V{k}_i)
\left(\frac{1}{k_1^3k_2^3}+\text{2 permutations}
\right)
\eea
The momentum shape is thus
\bea
B(\vec{k}_1 ,\vec{k}_2 , \vec{k}_3) &=& {1\over k_1^3 k_2^3} + {1\over k_1^3 k_3^3} +{1\over k_2^3 k_3^3}
\eea
which is entirely determined by the local shape ansatz.

\begin{figure}[h]
\centering
\includegraphics[width=2.5in]{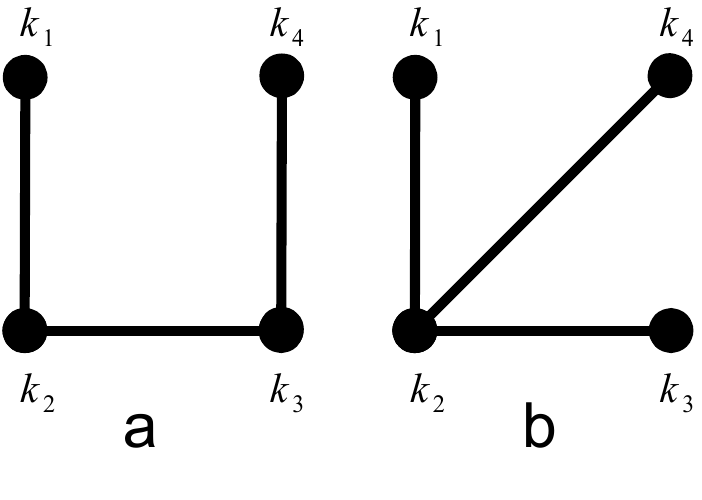}
\caption{Examples of tree-level contributions to $g_{NL}$ and $\tau_{NL}$.}
\label{GNLTNLtree}
\end{figure}

For the trispectrum there are two tree-level momentum shapes which are possible, because
there are two different structures for a diagram connecting 4 vertices with three
Gaussian correlators.  The first structure, which appears also at quadratic order
in the non-linear expansion for $\zeta$, arises from twelve diagrams like Fig.~\ref{GNLTNLtree}a and yields
the correlator terms
\bea
\langle \zeta_{\V{k}_1} \zeta_{\V{k}_2} \zeta_{\V{k}_3} \zeta_{\V{k}_4} \rangle
&=& (2\pi)^3 \delta^3 (\Sigma_i \vec{k}_i) A_2^2A_1^2 (2\pi^2\P)^3
\left( \frac{1}{k_1^3k_{12}^3k_4^3} +\text{11 permutations}\right)
\eea
where $k_{ij}= |\vec{k}_i + \vec{k}_j|$.
The other shape appears only at cubic order in the expansion, and is
given by four diagrams like Fig.~\ref{GNLTNLtree}b.  These diagrams yield correlator terms
\bea
\langle \zeta_{\V{k}_1} \zeta_{\V{k}_2} \zeta_{\V{k}_3} \zeta_{\V{k}_4} \rangle
&=& (2\pi)^3 \delta^3 (\Sigma_i \vec{k}_i) A_3 A_1^3 (2\pi^2\P)^3
\left( \frac{1}{k_1^3k_3^3k_4^3} +\text{3 permutations}\right)
\eea
Note that each tree-level diagram contributes to a term with a different momentum dependence.

We can define the 4-point momentum shapes
\bea
T(\vec{k}_1 ,\vec{k}_2, \vec{k}_3, \vec{k}_4) &=&  \frac{1}{k_1^3k_{12}^3k_4^3} +\text{11 permutations}
\nonumber\\
G(\vec{k}_1 ,\vec{k}_2, \vec{k}_3, \vec{k}_4) &=& \frac{1}{k_1^3k_3^3k_4^3} +\text{3 permutations}
\eea
If we write these tree-level correlators in the form
\bea
\label{correlators}
\langle \zeta_{\vec{k}_1} \zeta_{\vec{k}_2} \rangle &=&
(2\pi)^3 \delta^3 (\vec{k}_1+\vec{k}_2) (2\pi^2\mathcal{P})
F(\vec{k}_1 , \vec{k}_2) \times N^2
\nonumber\\
\langle \zeta_{\vec{k}_1} \zeta_{\vec{k}_2} \zeta_{\vec{k}_3} \rangle
&= & (2\pi)^3 \delta^3(\Sigma_i\V{k}_i) (2\pi^2\mathcal{P})^2
B(\vec{k}_1 ,\vec{k}_2 , \vec{k}_3) \times f
\nonumber\\
\langle \zeta_{\V{k}_1} \zeta_{\V{k}_2} \zeta_{\V{k}_3} \zeta_{\V{k}_4} \rangle
&=& (2\pi)^3 \delta^3 (\Sigma_i \vec{k}_i)  (2\pi^2\P)^3
T(\vec{k}_1 ,\vec{k}_2, \vec{k}_3, \vec{k}_4) \times t
\nonumber\\
&\,& +(2\pi)^3 \delta^3 (\Sigma_i \vec{k}_i) (2\pi^2\P)^3
G(\vec{k}_1 ,\vec{k}_2, \vec{k}_3, \vec{k}_4) \times g
\eea
then we can parameterize non-Gaussianity with the simple definitions
\bea
f_{NL} &\equiv& {5\over 6} {f \over N^4}
\nonumber\\
\tau_{NL} &\equiv& {t \over N^6}
\nonumber\\
g_{NL} &\equiv&  {g \over N^6}\;.
\eea
From our tree-level result
\bea
N^2 &=& C_1^2 + A_1^2 > A_1^2
\nonumber\\
f &=& A_2 A_1 ^2
\nonumber\\
t &=& A_2^2 A_1^2
\nonumber\\
g &=& A_3 A_1^3
\eea
it is clear that the constraint $\tau_{NL} \geq [(6/5)f_{NL}]^2$ holds at tree-level.

\section{Calculating Loop Diagrams}

However, loop diagrams can correct the correlators in eq.~\ref{correlators}, and can do so
in a scale-dependent manner.
In this section we outline a procedure for evaluating loop contributions to
$n$-point curvature perturbation correlators.

\begin{figure}[h]
\centering
\includegraphics[width=5in]{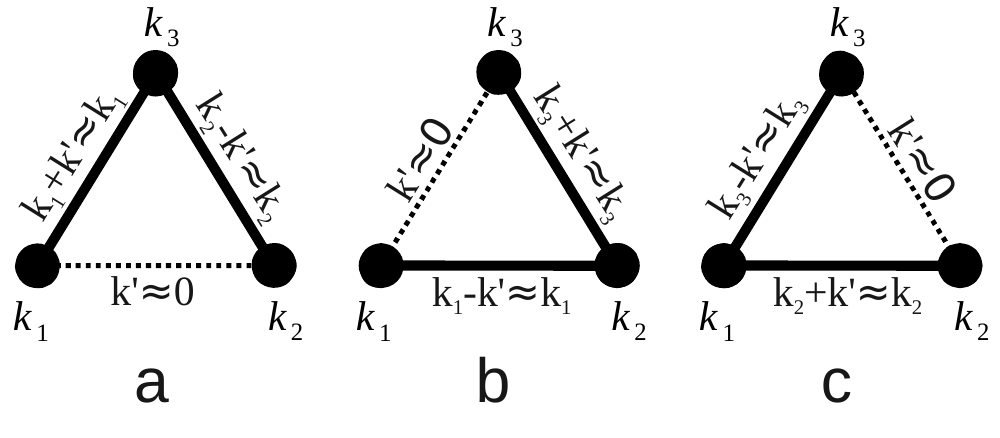}
\caption{One-loop corrections to the bispectrum up to quadratic order in the expansion.}
\label{FNL1loop}
\end{figure}

There is a single one-loop contribution to the bispectrum at quadratic order in the
local expansion, yielding the integral
\bea
I &=&\int_{1/L}^{\infty} \frac{d^3 \V{k}'}{(2\pi)^3}
{(2\pi^2 {\cal P})^3 \over k'^3|\vec{k}_1 -\vec{k}'|^3 |\vec{k}_2 +\vec{k}'|^3}\;.
\label{poleintegral}
\eea
This integral can be well approximated~\cite{Kumar:2009ge} by evaluating the integral near
the leading logarithmic singularities:
${\vec{k}' \sim 0}$, ${\vec{k}' \sim \vec{k}_1}$ and ${\vec{k}' \sim -\vec{k}_2}$.  Each
singularity corresponds to the momentum of one of the three correlators in the loop diagram becoming small.
When evaluating a diagram in the leading approximation, we will denote such a correlator with
a dashed line (see Fig.~\ref{FNL1loop}).
The resulting expression
\bea
I &\approx & (2\pi^2 {\cal P})^2 {\cal P}\left[\frac{\ln({\min}(k_1,k_2) L )}{ k_1^3k_2^3}
+ \frac{\ln({\min}(k_1,k_3) L )}{ k_1^3k_3^3} +\frac{\ln({\min}(k_2,k_3) L )}{ k_2^3k_3^3}
\right]
\eea
captures the leading behavior of this diagram.  It is easy to see why the behavior above
arises: the integral near each singularity is logarithmic, with one correlator momentum small
and the other two correlator momenta being approximately equal to external momenta.  The lower limit
of integration is set by the IR cutoff $L^{-1}$, while the upper limit is approximately the scale
at which at least one other (non-singular) correlator denominator starts to grow with the loop momentum.
Above this scale, the denominator will grow as $k_{loop}^6$ or greater and the integrand will become
small quickly.

More generally, one can derive a prescription for evaluating the leading behavior of an
arbitrary $l$-loop contribution to an $n$-point diagram.  The leading behavior of the integral is
dominated by the limit where the denominators of $l$ of the internal correlators become small.  If we denote these
correlators by dashed lines, it is clear that in the limit where the momenta associated with the dashed
lines vanishes, the remaining solid lines represent the correlators of a tree-level diagram with momenta
which are roughly linear combinations of the external insertion momenta.
As long as the momenta of the dashed-line correlators are small, the momenta of the solid-line correlators
only differ from the tree-level value by a small amount, which has an insignificant effect on those
correlators.  We may factor out this tree-level
contribution, and the leading loop contribution corresponds to a tree-level diagram, multiplied by
a loop integral of the form
\bea
\left({A_i A_j \over A_{i-1} A_{j-1}} \right)
\int_{1/L}^{k_{max_1}} {d^3 k_{loop,1} \over (2\pi)^3} {(2\pi^2) {\cal P} \over k_{loop,1}^3}
\times
\left({A_k A_l \over A_{k-1} A_{l-1}} \right)
\int_{1/L}^{k_{max_2}} {d^3 k_{loop,2} \over (2\pi)^3} {(2\pi^2) {\cal P} \over k_{loop,2}^3}
\nonumber\\
\times \ldots
\\ = \left({A_i A_j \over A_{i-1} A_{j-1}} \right){\cal P} \ln(k_{max,1}L)
\times \left({A_k A_l \over A_{k-1} A_{l-1}} \right) {\cal P} \ln(k_{max,2}L)
\nonumber\\
\times \ldots
\eea
where the $k_{loop,i}$ are the momenta of the dashed-line correlators.
Each dashed-line correlator is part of exactly one loop which contains
only that dashed-line correlator and other solid-line
correlators\footnote{If there existed two distinct loops
containing the same dashed-line correlator, then those
two together would form a loop of only solid-line correlators, violating the constraint that the solid-line
correlators form a tree-diagram.};
we denote by $\vec{q}_j$ the momenta of the solid-line correlators in
that loop, in the limit where all dashed-line correlator momenta vanish
(so the solid-line correlators form a tree-diagram, and the $\vec{q}_j$
are linear combinations of the external momenta).
We then define $k_{max} = \min(q_j)$ for each dashed-line correlator.
As long as $k_{loop,i}< k_{max,i}$ the momenta of the solid-line correlators are
largely independent of $k_{loop,i}$ and can be factored out of the integral.
For $k_{loop,i} > k_{max,i}$, at least one of the solid-line correlators
has a denominator which starts to grow with $k_{loop,i}$.
The integrand then scales as $k_{loop,i}^{-6}$ and is small
enough that we are justified in ignoring this region of the domain.
The factors $A_i A_j / A_{i-1} A_{j-1}$ account for different vertex coefficients for a loop-diagram,
as opposed to a tree-diagram.

One might worry that in a complex multi-loop diagram, the loop integrals would not factorize.
But it is straightforward to convince oneself that this is not the case.  The leading logarithmic
behavior of the integral is dominated by the region where all loop momenta are small, so
the integration limits can have only a small dependence on the loop momenta. More physically,
one might worry that the correlator denominators might remain small even as the loop momenta
became large if two loop momenta cancelled each other in a correlator denominator.  But the
logarithmic behavior depends on integration over the full loop momentum phase space as one
moves away from a singular point; if one restricts the loop momentum phase space by requiring
two momenta to cancel each other, then the remaining phase space numerator is insufficient to
cancel the $\Pi_i k_{loop,i}^{-3}$ behavior of the correlators, and the integrand still becomes
small.

We have checked this prescription for the leading logarithmic approximation by
comparing it to numerical integrations of the exact integrand for a few specific choices
of the external momentum insertions.  As in \cite{Kumar:2009ge} (Appendix B), we transformed
momentum integrals to the basis $\vec{n} = \vec{k}L$ and imposed the IR cutoff by
setting the integrand to zero when $|\vec{n}-\vec{n}_{pole}| \leq 1$.
Choosing $|\V{k_{small}}L|\approx 10$ and $|\V{k_{big}}L|\approx 1000$,
our approximation of one-loop corrections to squeezed bispectrum and trispectrum differed
from numerical values by $\lesssim 1\%$   The leading approximation to the two-loop
correction to the bispectrum typically differed from the numerical calculation by $\lesssim 5\%$
($\lesssim 10\%$ for the trispectrum).

We have shown that this approximation yields the leading behavior of the one-loop 3-point diagram
described above.
Extending this approximation to $l$-loop $n$-point diagrams, we arrive at the following rules:
\begin{enumerate}
\item
The numerator for each $n$-point diagram includes
\be
A_aA_b...A_z(2\pi^2 {\cal P})^{n-1} \P^l
(2\pi)^3 \delta^3(\Sigma_i \V{k}_i)
\ee
where the coefficients $A_a...$ correspond to the coefficients of the $n$ vertices outlined previously.
\item
To write the leading log approximation to the diagram, we identify the leading singularities.
There are $n+l-1$ Gaussian correlators, of which, in the leading approximation, $n-1$ can
be written as solid lines which form a tree-level diagram.  The remaining $l$ loop correlators have small
momenta and can be written as dashed lines (every choice of which lines are written as dashed corresponds to
a different term in the leading approximation).
The denominator of each $n$-point diagram includes a cubed product of tree-level momenta.
\be
{1 \over |\V{q}_1|^3...|\V{q}_{n-1}|^3}
\ee
where the momenta $q_i$ are only functions of the external momenta.  This is one term in one
of the local momentum shapes for the $n$-point function.
\item
Each dashed correlator spans a unique loop containing $m$ tree-level momentum lines $(\V{q}_1, ..., \V{q}_m)$ and
no other dashed-line correlators.  The momenta $\vec{q}_i$ are thus linear combinations of the
external momentum insertions, $\vec{k}_i$.
The integral over each dashed correlator momentum yields a factor
\be
\ln{ (\min(\V{q}_1, ..., \V{q}_m)L)}\;.
\ee
Note that in the case where a dashed correlator spans only one tree-level line $\V{k}_i$, this simply evaluates to $\ln{ (\V{k}_i L)}$.
\item
If a set of $r$ dashed lines and $s$ solid lines ($s=0,1$) start at the same vertex and end at the same vertex,
this results in a
symmetry factor of ${1 \over r!}$ if $s=0$, or ${1 \over (r-1)!}$ if $s=1$. (Note that while
calculating four-point diagrams at the cubic order, $r \leq 2$.)
\end{enumerate}

We see from these rules that each loop comes with a factor of the form
\bea
\text{``Loop factor"} &=& {A_i A_j \over A_{i-1} A_{j-1}} {\cal P}\ln(kL)
\eea
In general, one might expect these factors to be small, since ${\cal P} = (H / 2\pi )^2$ is necessarily
small.  However, simple models can be constructed where these loop factors are not small and can even
dominate over the tree-level term~\cite{Kumar:2009ge}.

\section{Results}

We can now apply this procedure to the calculation of the power spectrum, bispectrum and trispectrum
for local ansatz, to cubic order in the expansion and to 3-loop order.

\subsection{Power Spectrum}

\begin{figure}[h]
\centering
\includegraphics[width=5in]{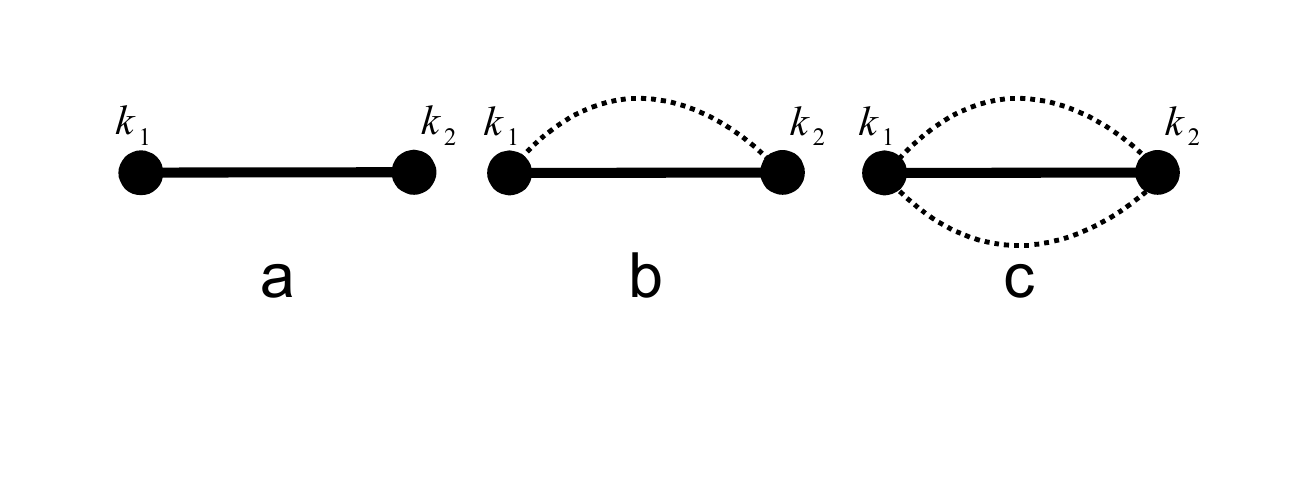}
\caption{Tree-level, one-loop and two-loop diagrams contributing to the power spectrum.}\label{powerspectrum}
\end{figure}

All diagrams\footnote{There are also loop diagrams with dressed vertices, but we will assume that those 
contributions have already been absorbed into the corrected vertex coefficient.} 
contributing to the power spectrum in the cubic expansion are given in Fig.~\ref{powerspectrum}.
The full expression for the power spectrum (up to two loop order) is given by
\bea
\Expect{\zeta_{\V{k}_1}\zeta_{\V{k}_2}}
&=& (2\pi)^3 \delta^3(\V{k_1}+\V{k}_2)F(\vec{k}_1, \vec{k}_2) (2\pi^2 {\cal P}_\zeta)
\nonumber\\
&=& (2\pi)^3 \delta^3(\V{k_1}+\V{k}_2)F(\vec{k}_1, \vec{k}_2) (2\pi^2 {\cal P})
\left[C^2 + A_1^2 + A_2^2\mathcal{P}\ln(k_1L) +  \frac{A_3^2}{2}\mathcal{P}^2\ln(k_1L)^2\right]
\nonumber\\
\eea
Defining $n_s = d\ln {\cal P}_\zeta / d\ln k $, we find
\bea
n_s &=& {A_2^2\mathcal{P} +  A_3^2\mathcal{P}^2\ln(kL) \over
C^2 + A_1^2 + A_2^2\mathcal{P}\ln(kL) +  \frac{A_3^2}{2}\mathcal{P}^2\ln(kL)^2}
\eea
COBE measurements of the power spectrum normalization and WMAP bounds on the running
of the power spectrum~\cite{Komatsu:2008hk} give us the rough constraints
\bea
\label{powerspectrumconstraint}
N^2 {\cal P} & \sim & 10^{-10}
\nonumber\\
{ A_2^2\mathcal{P} \over N^2} &\lesssim & 10^{-2}
\nonumber\\
{A_3^2\mathcal{P}^2\ln(kL)  \over N^2} &\lesssim & 10^{-2}
\eea

\subsection{Bispectrum}

Going beyond tree-level, in general one may write the 3-point correlator as
\bea
\langle \zeta_{\V{k}_1} \zeta_{\V{k}_2} \zeta_{\V{k}_3} \rangle &=& (2\pi)^3
\delta^3 (\V{k}_1 + \V{k}_2 +\V{k}_3) \langle \zeta_{\V{k}_1} \zeta_{\V{k}_2} \zeta_{\V{k}_3} \rangle '
\eea
and one may then define $f_{NL}$ in the squeezed-limit (see,
for example, ~\cite{Smith:2006ud,Smith:2011if}) as
\bea
f_{NL}^{sq.} &=& {5 \over 12} \lim_{\V{k}_1 \rightarrow 0}
{\langle \zeta_{\V{k}_1} \zeta_{\V{k}_2} \zeta_{\V{k}_3} \rangle ' \over
P_\zeta (\V{k}_1) P_\zeta (\V{k}_2) }
\eea

The full 3-point correlator to third order in $\chi$ is computed in the Appendix, and is
given by the expression
\bea
\langle \zeta_{\V{k}_1} \zeta_{\V{k}_2} \zeta_{\V{k}_3} \rangle &=& A_2(2\pi^2\P)^2 (2\pi)^3 \delta^3(\Sigma_i\V{k}_i){ 1 \over k_2^3k_1^3 }
\bigg[A_1^2 + A_2^2\P\ln{({\min(k_1,k_2)L)}}+ A_1A_3\P\ln{{(k_1k_2L^2)}}
\nonumber \\
&+& A_3^2\P^2\ln{({\min(k_1,k_2)L)}}\left(\ln{{(k_1k_2L^2)}}+{1\over 2}\ln{({\min(k_1,k_2)L)}}\right) \bigg]
+ 2\;\rm{perms}
\eea
It is illuminating to consider this correlator in the squeezed limit $k_{big}\sim k_2 \sim k_3 \gg k_1 \sim k_{sm.}$:
\bea
\langle \zeta_{\V{k}_1} \zeta_{\V{k}_2} \zeta_{\V{k}_3} \rangle &\sim& A_2 A_1^2 (2\pi^2\P)^2 (2\pi)^3
\delta^3(\Sigma_i\V{k}_i){ 1 \over k_2^3k_1^3 }
\left[1 + {A_2^2 \over A_1^2} \right. \P\ln{({k_1 L})}+ {A_3 \over A_1}\P\ln{{(k_2L)}}
\nonumber \\
&+& \left. {A_3^2 \over A_1^2}\P^2\ln{(k_1L)}\ln{(k_2L)} \right]
%+ 2\;\rm{perms}
+ (\vec{k}_2 \leftrightarrow \vec{k}_3)
\eea
In the squeezed limit, the coefficient $f_{NL}$ is then given by
\bea
f_{NL}^{sq.} &=& {5\over 6} {A_2 A_1^2 \over N^4}
\left[1 + {A_2^2 \over A_1^2}\P\ln{({k_{sm.} L})}+ {A_3 \over A_1}\P\ln{{(k_{big}L)}}
+ {A_3^2 \over A_1^2}\P^2\ln{(k_{sm.}L)}\ln{(k_{big}L)}\right],
\nonumber\\
\eea
while in the equilateral limit ($k_1 \sim k_2 \sim k_3$) we instead get
\bea
f_{NL}^{equi.} &=& {5\over 6} {A_2 A_1^2 \over N^4}
\left[1 + {A_2^2 \over A_1^2}\P\ln({kL})+ 2{A_3 \over A_1}\P\ln(kL)
+ {5\over 2} {A_3^2 \over A_1^2} \P^2 (\ln(kL))^2\right].
\nonumber\\
\eea
The expression for $f_{NL}^{equi.}$ is consistent with the one-loop
expression given in~\cite{Sugiyama:2011jt}.

Note that, while in the equilateral limit the loop contribution is always of the
same sign as the tree-level contribution, this is not the case for the
squeezed limit.  If $\ln (k_{sm.}L) \sim 0$, then
\bea
f_{NL}^{sq.} &\rightarrow & {5\over 6} {A_2 A_1^2 \over N^4}
\left[1 +  {A_3 \over A_1}\P\ln{{(k_{big}L)}}\right]
\eea
and the one-loop contribution dominates if $|(A_3 / A_1) {\cal P} \ln (k_{big}L)| > 1$.
If the loop term dominates, then the power spectrum constraints (Eq.~\ref{powerspectrumconstraint})
from COBE and WMAP imply
$|f_{NL}^{sq.}| \lesssim 800 (\ln k_{big}L)^{1\over 2} $.
Note that these constraints would permit a one-loop contribution to the $f_{NL}$ in the
squeezed limit which is roughly an order of magnitude larger than that permitted at
quadratic order in the equilateral limit~\cite{Kumar:2009ge}.
Of course, direct constraints
from WMAP on $f_{NL}$ are considerably tighter, requiring
$-4 < f_{NL}^{sq.}<80$~\cite{Smith:2009jr}.

It is especially interesting to note that, if $A_3 / A_1$ is negative and
$|(A_3 / A_1) {\cal P} \ln (k_{big}L)| \sim 1$, then the sign of
$f_{NL}^{sq.}$ can change as $k_{big}$ is varied.  This could have an especially
interesting effect on galaxy
bias~\cite{galaxybias}.

\subsection{Trispectrum}

The full trispectrum to all orders in the cubic expansion (three-loop) is computed
in Appendix B (as the full expression is quite large, it will not be reproduced
here).  We will instead consider the simplest limit, where all the momenta
are assumed to be of roughly the same scale ($k_i \sim k_{ij} \sim k$).  In this limit, we can write
\bea
\tau_{NL}^{equi.} &=& {A_1^2 A_2^2 \over N^6} \left[1+  \left({A_2^2 \over A_1^2}
+ 4{A_3 \over A_1} + {A_3^2 \over A_2^2}\right) \right. {\cal P}\ln(kL)
+\left({15 \over 2}{A_3^2 \over A_1^2} + 2{A_3^3 \over A_2^2 A_1}\right)({\cal P}\ln(kL))^2
\nonumber\\ &\,&
+\left. {5\over 2}{A_3^4 \over A_2^2 A_1^2}({\cal P}\ln(kL))^3 \right]
\nonumber\\
g_{NL}^{equi.} &=& {A_3 A_1^3 \over N^6} \left[1+3{A_2^2 \over A_1^2}{\cal P}\ln(kL)
+\left(3{A_3 A_2^2 \over A_1^3}+{3\over 2}{A_3^2 \over A_1^2}\right)({\cal P}\ln(kL))^2
+{A_3^3 \over A_1^3}({\cal P}\ln(kL))^3\right]
\nonumber\\
\eea
The expression for $\tau_{NL}^{equi.}$ is consistent with the one-loop expression
given in~\cite{Sugiyama:2011jt}.

It is again illuminating to consider the behavior of the
trispectrum in the limit where the local trispectrum shape $T(\vec{k}_1 ,\vec{k}_2, \vec{k}_3, \vec{k}_4) $
becomes large, namely, the elongated quadrilateral limit:
$\vec{k}_1 \sim -\vec{k}_2 = \vec{k}_A$, $\vec{k}_3 \sim -\vec{k}_4 = \vec{k}_B$,
$\vec{k}_1 + \vec{k}_2 = \vec{k}_{sum}$, $|\vec{k}_A| \geq |\vec{k}_B| \gg |\vec{k}_{sum}|$.  In this limit
\bea
G(\vec{k}_1 ,\vec{k}_2, \vec{k}_3, \vec{k}_4) &\sim& {2\over k_A^3 k_B^6}
\nonumber\\
T(\vec{k}_1 ,\vec{k}_2, \vec{k}_3, \vec{k}_4) &\sim& {4\over k_A^3 k_{sum}^3 k_B^3}
\eea
We then find
\bea
\langle \zeta_{\V{k}_1} \zeta_{\V{k}_2} \zeta_{\V{k}_3} \zeta_{\V{k}_4} \rangle &\sim&
(2\pi)^3 \delta^3(\Sigma_i \vec{k}_i) (2\pi^2 {\cal P})^3 A_3 A_1^3 {1\over k_1^3 k_3^3 k_4^3}
\left[1+3{A_2^2 \over A_1^2} \right.  {\cal P} \ln (k_B L)
\nonumber\\
&\,&
+ 3\left({A_3 A_2^2 \over A_1^3 }+{1\over 2}{A_3^2 \over A_1^2} \right){\cal P}^2(\ln(k_B L))^2
+\left. {A_3^3 \over A_1^3}{\cal P}^3(\ln(k_B L))^3 \right]
\nonumber\\
&+& (2\pi)^3 \delta^3(\Sigma_i \vec{k}_i) (2\pi^2 {\cal P})^3 A_2^2 A_1^2 {1\over k_1^3 k_{12}^3 k_4^3}
\left[1+ \left({A_2^2 \over A_1^2}+{A_3^2 \over A_2^2}\right) \right. {\cal P} \ln(k_{sum}L)
\nonumber\\ &\,&
+ {A_3 \over A_1}{\cal P} [\ln((k_A)L)+\ln((k_B)L)]
+{A_3^2 \over A_1^2}{\cal P}^2 \ln(k_A L)\ln(k_B L)
\nonumber\\ &\,&
+{A_3^3 \over A_2^2 A_1}{\cal P}^2 [\ln((k_A)L)+\ln((k_B)L)]\ln(k_{sum}L)
\nonumber\\ &\,&
+\left. {A_3^4 \over A_2^2 A_1^2}{\cal P}^3 (\ln(k_B L))^2\ln(k_{sum}L)
\right] +\text{permutations}
\eea
where the first two lines contribute to $g_{NL}$ and the last four lines contribute
to $\tau_{NL}$.

If we take the limit $\ln(k_{sum}L) \sim 0$, then we find
\bea
\tau_{NL}^{sq.} \rightarrow {A_2^2 A_1^2 \over N^6} \left[1+{A_3 \over A_1}{\cal P}\ln(k_A L) \right]
 \left[1+{A_3 \over A_1}{\cal P}\ln(k_B L) \right]
\eea
with the dominant loop contributions arising from diagrams ``$\tau_{NL}$-e", ``$\tau_{NL}$-g", and
``$\tau_{NL}$-m" in Appendix B.2.  If we take $k_A = k_B \equiv k_{big}$, then
the constraint $\tau_{NL} \geq [(6/5)f_{NL}^{sq.}]^2$ is necessarily satisfied; at the
scale $k_{big}$ where $f_{NL}$ changes sign, $\tau_{NL}$ also vanishes because loop corrections
cancel the tree-level contribution.

But if we relax this constraint, allowing $k_B < k_A$, then $\tau_{NL}$ can be negative.
In particular, in the limit $k_A = k_{big}$ and ${A_3 \over A_1}{\cal P}\ln(k_B L),
{A_3 \over A_1}{\cal P} \ln(k_{sum}L) \sim 0$, we find
\bea
\tau_{NL}^{sq.} \sim {A_2^2 A_1^2 \over N^6 } \left[1+ {A_3 \over A_1}{\cal P} \ln(k_{big}L) \right]
\eea
where the dominant loop contribution arises from diagram ``$\tau_{NL}$-g" in the Appendix B.2.
As with the squeezed limit of $f_{NL}$, the one-loop contribution to
$\tau_{NL}^{sq.}$ will dominate over the tree-level contribution if
$|(A_3 / A_1) {\cal P} \ln(k_{big}L)| >1$.  The sign of the loop-contribution can be
positive or negative, depending on the relative sign of $A_1$ and $A_3$.
As a result, $\tau_{NL}$ could indeed be negative.

This result differs from that stated in~\cite{Sugiyama:2011jt}, because that work
assumed that all external momenta are of the same order.  We can compare our result
to that in~\cite{Smith:2011if}, where it was claimed that the relation
$\tau_{NL} \geq [(6/5)f_{NL}^{sq.}]^2$ is always satisfied, due to the fact that
a particular covariance matrix is always positive definite.  This argument relies
on the implicit assumption $k_A = k_B$ made in~\cite{Smith:2011if}.  As we have seen above,
we reproduce this result if this assumption is made.  But this assumption is not
required.

Constraints on the power spectrum from COBE and WMAP (Eq.~\ref{powerspectrumconstraint}) constrain the loop contribution
to $\tau_{NL}$ to have a magnitude less than $\sim 10^7 \sqrt{\ln (k_{big}L)}$.  As with
$f_{NL}$ these constraints in the squeezed limit are much less tight than those at quadratic
order in the equilateral limit~\cite{Kumar:2009ge}.  Indeed, direct constraints from WMAP already
constrain $|\tau_{NL}| \lesssim 10^4$ ~\cite{Smidt:2010zy}.

\section{Conclusions}

In this work, we have considered the local ansatz for non-Gaussianity in curvature perturbations.
We have introduced a simple general formalism for computing higher-order loop corrections to
general $n$-point correlators.  As expected, the general leading behavior of loop diagrams is
the same as for tree-level diagrams, up to logarithmic corrections.

We have illustrated this procedure by computing the curvature power spectrum, bispectrum and trispectrum
(expanded to cubic order in fundamental Gaussian fields) to 3-loop order.  This calculation has
been made for general choices of the external momentum insertions.
In particular we conclude that the relation
$\tau_{NL} \geq (1/2)[(6/5) f_{NL}]^2$ is not necessarily obeyed for all momentum shapes.  Furthermore,
there exists a squeezed limit (where the $\tau_{NL}$ momentum structure is dominant) where $\tau_{NL}$ can be
negative if the loop-contribution dominates.  It has also been found that in the squeezed limit
the sign of $f_{NL}^{sq.}$ can change as a function of scale.  This can have an interesting effect
on galaxy bias.

Note that the relationship between $f_{NL}$ and $\tau_{NL}$ depends upon the relative sizes of the external momenta.
The constraint $\tau_{NL} \geq [(6/5) f_{NL}^{sq.}]^2$ is obtained in the elongated equilateral limit if the
external momenta of the trispectrum are taken to all be of the same magnitude, while $\tau_{NL} <0$ can be obtained
if there is even a modest hierarchy in the external momenta.  The relation $\tau_{NL} \geq (1/2)[(6/5) f_{NL}]^2$ is only
obeyed at one-loop order if the external momenta are taken to be
of the same size.  It is important to remember that even if non-Gaussianities are in fact generated by
the exact local ansatz (assuming the $A_i$ are constant), the bispectrum and trispectrum will not have an
exactly local shape; each term in the local shape will be scaled by logarithmic corrections which depend
on the particular choice of external momenta.  Unless the loop contributions are negligible, the determination
of parameters such as $f_{NL}$, $\tau_{NL}$ and $g_{NL}$ from the data will depend on what estimators are used
and to which regions of momentum space they are sensitive~\cite{Kamionkowski:2010me}.
It would be interesting to determine what values of
these non-Gaussian parameters would actually be yielded by standard estimators as a function of the coefficients
$A_i$ of the local ansatz.
\\
{\bf Acknowledgements}
\\
We thank E.~Komatsu, L.~Leblond, A.~Rajaraman and S.~Shandera for useful discussions.
Preliminary results of this work have been presented at the Aspects Of Inflation Workshop
and at the Cosmological non-Gaussianity: Observation Confronts Theory
Workshop.  We are grateful to the organizers of these workshops, Texas A\&M University (MIFP)
and the University of Michigan (MCTP) for their hospitality.
The work of JB and JK was supported in part by the Department of Energy under
Grant~DE-FG02-04ER41291.

\pagebreak
\appendix

\section{Bispectrum}

\begin{figure}[h]
\centering
\includegraphics[width=6in]{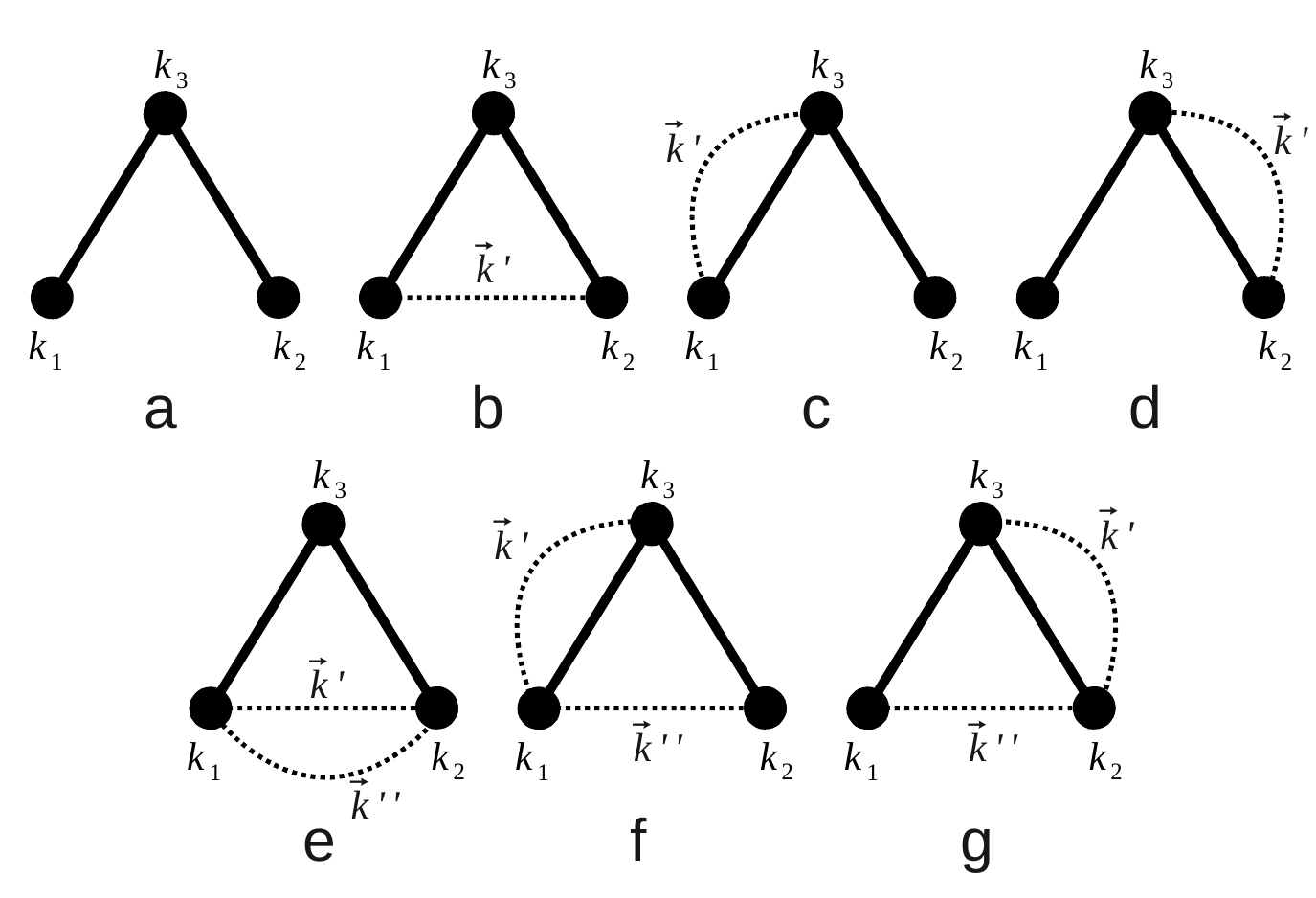}
\caption{The bispectrum and all distinct corrections up to third order are displayed. Note that particular momentum labels have
 been chosen for clarity, while for the purpose of calculation all possible momenta permutations are included.}
\label{FNL}
\end{figure}

Computing the bispectrum to two loops in the squeezed limit requires a careful examination of the momentum
structure of corresponding IR divergent integrals.
In Fig.~\ref{FNL}, we exhibit all diagrams contributing to bispectrum terms with leading
momentum behavior $ 1 / k_1^3 k_2^3$.
All other bispectrum contributions are related to this result by permutation of the external momenta.

The tree-level contribution to the bispectrum is given by
\bea
\langle \zeta_{\V{k}_1} \zeta_{\V{k}_2} \zeta_{\V{k}_3} \rangle_{f_{NL}a}
&= & \frac{A_1^2A_2(2\pi^2\mathcal{P})^2}{2}(2\pi)^6
\int{\frac{d^3\V{k'}}{(2\pi)^3}
\left( \frac{\delta^3(\V{k_1}+\V{k'}-\V{k_3})
\delta^3(\V{k_2}-\V{k'})}{(k_1k_2k'(k_3-k'))^{3/2}}+1\;\rm{variant} \right)}
\nonumber \\
&= & \frac{2A_1^2A_2(2\pi^2\mathcal{P})^2}{2}(2\pi)^3
\delta^3(\Sigma_i\V{k_i})
\frac{1}{k_1^3k_2^3}
\nonumber \\
&= & A_1^2A_2(2\pi^2\mathcal{P})^2(2\pi)^3
\delta^3(\Sigma_i\V{k_i})
\frac{1}{k_1^3k_2^3}
\eea
The one-loop correction to $f_{NL}$ which uses only terms to second order in $\chi$ is
\bea
\langle \zeta_{\V{k}_1} \zeta_{\V{k}_2} \zeta_{\V{k}_3} \rangle_{f_{NL}b}
&=& \frac{A_2^3}{8} \int \frac{d^3 \V{k}'d^3 \V{k}''d^3 \V{k}'''}{(2\pi)^9}
\langle (\chi_{\V{k}_1-\V{k}'} \chi_{\V{k}'})
(\chi_{\V{k}_2-\V{k}''} \chi_{\V{k}''})
(\chi_{\V{k}_3-\V{k}'''} \chi_{\V{k}'''})\rangle\; ,
\nonumber\\
&=& \frac{A_2^3(2\pi^2 \P)^3}{8}
(2\pi)^3 \delta^3(\Sigma_i\V{k_i})
\int{d^3 \V{k}' \over (2\pi)^3 } \left({1
\over k'^3
|\vec{k_1} -\vec{k'}|^3 |\vec{k_2} +\vec{k'}|^3 } + 7 \; \rm{perms}\right)
\nonumber \\
&=& A_2^3(2\pi^2)^2 \P^3
(2\pi)^3 \delta^3(\Sigma_i\V{k_i})
\frac{\ln ({\min}(k_1,k_2) L)}{k_1^3k_2^3} \; .
\eea
The \ref{FNL}c piece of the bispectrum includes an integral with two poles instead of three:
\bea
\langle \zeta_{\V{k}_1} \zeta_{\V{k}_2} \zeta_{\V{k}_3} \rangle_{f_{NL}c}
&=& {1\over2}A_3A_2A_1(2\pi^2 \P)^3
(2\pi)^3 \delta^3(\Sigma_i\V{k_i})
\int{d^3 \V{k}' \over (2\pi)^3 } \left({1
\over {k'^3} |\vec{k_1} -\vec{k'}|^3 {k_2^3}}\right)
\nonumber \\
& = &  A_3A_2A_1(2\pi^2)^2 \P^3
(2\pi)^3 \delta^3(\Sigma_i\V{k_i})
\frac{\ln (\rm k_1L)}{k_1^3k_2^3}
\eea
The \ref{FNL}d piece is a mirror of \ref{FNL}c.
\bea
\langle \zeta_{\V{k}_1} \zeta_{\V{k}_2} \zeta_{\V{k}_3} \rangle_{f_{NL}d}
&=&  {1\over2} A_3A_2A_1(2\pi^2 \P)^3
(2\pi)^3 \delta^3(\Sigma_i\V{k_i})
\int{d^3 \V{k}' \over (2\pi)^3 } \left({1
\over {k'^3} |\vec{k_2} -\vec{k'}|^3 {k_1^3}}\right)
\nonumber \\
& = &  A_3A_2A_1(2\pi^2)^2 \P^3
(2\pi)^3 \delta^3(\Sigma_i\V{k_i})
\frac{\ln (\rm k_2L)}{k_1^3k_2^3}
\eea
The  \ref{FNL}e,  \ref{FNL}f, and  \ref{FNL}g contributions share the same shape, but must be evaluated
separately because the integrals they produce are different.
For  \ref{FNL}e the expression again relies on which of $|k_1|$ and $|k_2|$ is smaller.
\bea
&\langle \zeta_{\V{k}_1} \zeta_{\V{k}_2} \zeta_{\V{k}_3} \rangle_{f_{NL}e}&
\nonumber\\
&=&{1 \over 2}A_3^2A_2(2\pi^2 \P)^4
(2\pi)^3 \delta^3(\Sigma_i\V{k_i})
\int \int{d^3 \V{k}'^3d^3\V{k}''^3 \over (2\pi)^6 } \left({1
\over {|\V{k_1}-\V{k'}+\V{k''}|^3 |\vec{k_2} +\vec{k'} -\vec{k''}|^3k'^3 k''^3}}\right)
\nonumber \\
&=&{1 \over 2}A_3^2A_2(2\pi^2)^3 \P^4
(2\pi)^3 \delta^3(\Sigma_i\V{k_i})
\int{d^3\V{k}'^3 \over (2\pi)^3} \left({\ln({{\min(k_1,k_2)L}})
\over {|\V{k_1}-\V{k'}|^3 |\vec{k_2} +\vec{k'}|^3k'^3}}\right)
\nonumber \\
&=&{1 \over 2}A_3^2A_2(2\pi^2)^2 \P^4
(2\pi)^3 \delta^3(\Sigma_i\V{k_i})
 \frac{(\ln{({\min(k_1,k_2)L}})^2}{k_1^3k_2^3}
\nonumber\\
\eea
The  \ref{FNL}f and  \ref{FNL}g contributions mirror each other.
\bea
\langle \zeta_{\V{k}_1} \zeta_{\V{k}_2} \zeta_{\V{k}_3} \rangle_{f_{NL}f}
&=& {1\over2} A_3^2A_2(2\pi^2 \P)^4
(2\pi)^3 \delta^3(\Sigma_i\V{k_i})
\int \int{d^3 \V{k}'^3d^3\V{k}''^3 \over (2\pi)^6 } \left({1
\over {|\V{k_2}-\V{k'}|^3 |\vec{k_1} +\vec{k'} -\vec{k''}|^3k'^3 k''^3}}\right)
\nonumber \\
&=& {1\over2}A_3^2A_2(2\pi^2)^3 \P^4
(2\pi)^3 \delta^3(\Sigma_i\V{k_i})
\int{d^3\V{k}'^3 \over (2\pi)^3} \left({2\ln({{k_1L}})
\over {|\V{k_2}-\V{k'}|^3 |\vec{k_1} +\vec{k'}|^3k'^3}}\right)
\nonumber \\
&=&A_3^2A_2(2\pi^2)^2 \P^4
(2\pi)^3 \delta^3(\Sigma_i\V{k_i})
 \frac{\ln({{k_1L}})\ln{({\min(k_1,k_2)L)}}}{k_1^3k_2^3}
\\
\langle \zeta_{\V{k}_1} \zeta_{\V{k}_2} \zeta_{\V{k}_3} \rangle_{f_{NL}g}
&=&  {1\over2}A_3^2A_2(2\pi^2 \P)^4
(2\pi)^3 \delta^3(\Sigma_i\V{k_i})
\int \int{d^3 \V{k}'^3d^3\V{k}''^3 \over (2\pi)^6 } \left({1
\over {|\V{k_1}-\V{k'}|^3 |\vec{k_2} +\vec{k'} -\vec{k''}|^3k'^3 k''^3}}\right)
\nonumber \\
&=& {1\over2}A_3^2A_2(2\pi^2)^3 \P^4
(2\pi)^3 \delta^3(\Sigma_i\V{k_i})
\int{d^3\V{k}'^3 \over (2\pi)^3} \left({2\ln({{k_2L}})
\over {|\V{k_1}-\V{k'}|^3 |\vec{k_2} +\vec{k'}|^3k'^3}}\right)
\nonumber \\
&=&A_3^2A_2(2\pi^2)^2 \P^4
(2\pi)^3 \delta^3(\Sigma_i\V{k_i})
 \frac{\ln({{k_2L}})\ln{({\min(k_1,k_2)L)}}}{k_2^3k_1^3}
\eea
Collecting all terms, the full $f_{NL}$ contribution to third order in $\chi$ is
\bea
\langle \zeta_{\V{k}_1} \zeta_{\V{k}_2} \zeta_{\V{k}_3} \rangle &=& A_2(2\pi^2\P)^2 (2\pi)^3 \delta^3(\Sigma_i\V{k_i}){ 1 \over k_2^3k_1^3 }
\bigg[A_1^2 + A_2^2\P\ln{({\min(k_1,k_2)L)}}+ A_1A_3\P\ln{{(k_1k_2L^2)}}
\nonumber \\
&+& A_3^2\P^2\ln{({\min(k_1,k_2)L)}}\left(\ln{{(k_1k_2L^2)}}+\left.
{1\over 2}\ln{({\min(k_1,k_2)L)}}\right) \right]
+ 2\;\rm{perms}
\eea

\section{Trispectrum}

In this appendix we list the complete momentum structure of 4-point correlators up to
cubic non-linear order in local expansion of $\zeta$ in Gaussian fields.  The complete
expansion contains diagrams up to three-loops.  Several of these diagrams will contain terms
(in the leading log approximation) which contribute to both $g_{NL}$ and $\tau_{NL}$ momentum
structures.  We will thus compute the contribution to each parameter separately.

\subsection{$g_{NL}$}
The integrals involved in computing $g_{NL}$ are nearly identical to those used for the bispectrum.
The resulting contributions from each diagram are listed below. To save space, henceforth $K \equiv  (2\pi)^3\delta^3(\Sigma_ik_i)$.
\begin{figure}[h]
\centering
\includegraphics[width=5.8in]{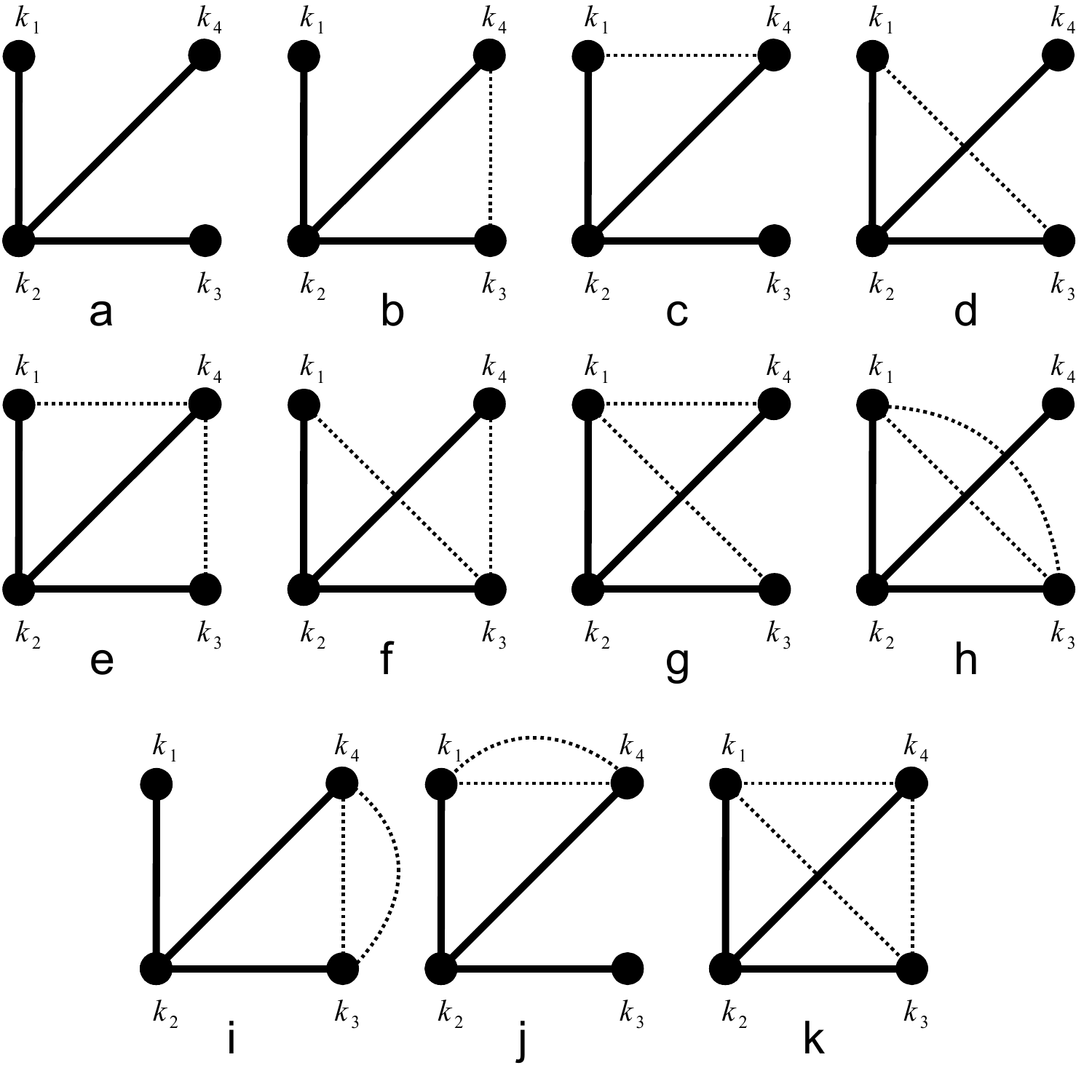}
\caption{Diagrams contributing to $g_{NL}$ at tree level and all corrections to third order in $\chi$.
These diagrams contribute to $1/k_1^3 k_3^3 k_4^3$ term of the 4-point correlator.}
\label{GNL}
\end{figure}
\bea
\langle \zeta_{\V{k}_1} \zeta_{\V{k}_2} \zeta_{\V{k}_3} \zeta_{\V{k}_4} \rangle_{g_{NL}a}
&=&A_3A_1^3K(2\pi^2\P)^3 \frac{1}{k_1^3k_4^3k_3^3}
\\
\langle \zeta_{\V{k}_1} \zeta_{\V{k}_2} \zeta_{\V{k}_3} \zeta_{\V{k}_4} \rangle_{g_{NL}b}
&=&A_3A_2^2A_1K(2\pi^2)^3\P^4 \frac{\ln{{(\min(k_3,k_4)L)}}}{k_1^3k_4^3k_3^3}
\\
\langle \zeta_{\V{k}_1} \zeta_{\V{k}_2} \zeta_{\V{k}_3} \zeta_{\V{k}_4} \rangle_{g_{NL}c}
&=&A_3A_2^2A_1K(2\pi^2)^3\P^4 \frac{\ln{{(\min(k_1,k_4)L)}}}{k_1^3k_4^3k_3^3}
\\
\langle \zeta_{\V{k}_1} \zeta_{\V{k}_2} \zeta_{\V{k}_3} \zeta_{\V{k}_4} \rangle_{g_{NL}d}
&=&A_3A_2^2A_1K(2\pi^2)^3\P^4 \frac{\ln{{(\min(k_1,k_3)L)}}}{k_1^3k_4^3k_3^3}
\eea
A new mathematical expression in these calculations comes from integrals of the form
\be
\int\int{d^3k'd^3k'' \over (2\pi)^6}
{1 \over {|\V{k_4}-\V{k'}+\V{k''}|^3 |\vec{k_1} +\vec{k'}|^3|\V{k_3}-\V{k''}|^3k'^3 k''^3}}
\approx \frac{\ln{{(\min(k_3,k_4)L)}}\ln{{(\min(k_1,k_4)L)}}}{k_1^3k_4^3k_3^3}
\label{gnlint}
\ee
\bea
\langle \zeta_{\V{k}_1} \zeta_{\V{k}_2} \zeta_{\V{k}_3} \zeta_{\V{k}_4} \rangle_{g_{NL}e}
&=&A_3^2A_2^2 K(2\pi^2)^3\P^5 \frac{\ln{{(\min(k_3,k_4)L)}}\ln{{(\min(k_1,k_4)L)}}}{k_1^3k_4^3k_3^3}
\\
\langle \zeta_{\V{k}_1} \zeta_{\V{k}_2} \zeta_{\V{k}_3} \zeta_{\V{k}_4} \rangle_{g_{NL}f}
&=&A_3^2A_2^2 K(2\pi^2)^3\P^5 \frac{\ln{{(\min(k_3,k_4)L)}}\ln{{(\min(k_1,k_3)L)}}}{k_1^3k_4^3k_3^3}
\\
\langle \zeta_{\V{k}_1} \zeta_{\V{k}_2} \zeta_{\V{k}_3} \zeta_{\V{k}_4} \rangle_{g_{NL}g}
&=&A_3^2A_2^2 K(2\pi^2)^3\P^5 \frac{\ln{{(\min(k_1,k_4)L)}}\ln{{(\min(k_1,k_3)L)}}}{k_1^3k_4^3k_3^3}
\eea
\bea
\langle \zeta_{\V{k}_1} \zeta_{\V{k}_2} \zeta_{\V{k}_3} \zeta_{\V{k}_4} \rangle_{g_{NL}h}
&=&{1 \over 2}A_3^3A_1 K(2\pi^2)^3\P^5 \frac{[\ln{{(\min(k_1,k_3)L)}}]^2}{k_1^3k_4^3k_3^3}
\\
\langle \zeta_{\V{k}_1} \zeta_{\V{k}_2} \zeta_{\V{k}_3} \zeta_{\V{k}_4} \rangle_{g_{NL}i}
&=&{1 \over 2}A_3^3A_1 K(2\pi^2)^3\P^5 \frac{[\ln{{(\min(k_3,k_4)L)}}]^2}{k_1^3k_4^3k_3^3}
\\
\langle \zeta_{\V{k}_1} \zeta_{\V{k}_2} \zeta_{\V{k}_3} \zeta_{\V{k}_4} \rangle_{g_{NL}j}
&=&{1 \over 2}A_3^3A_1 K(2\pi^2)^3\P^5 \frac{[\ln{{(\min(k_1,k_4)L)}}]^2}{k_1^3k_4^3k_3^3}
\eea
Finally, much as in Eq \ref{gnlint}, the fully connected four-point diagram evaluates to
\be
\langle \zeta_{\V{k}_1} \zeta_{\V{k}_2} \zeta_{\V{k}_3} \zeta_{\V{k}_4} \rangle_{g_{NL}k}
=A_3^4 K(2\pi^2)^3\P^6 \frac{\ln{{(\min(k_1,k_3)L)}}\ln{{(\min(k_3,k_4)L)}}\ln{{(\min(k_1,k_4)L)}}}{k_1^3k_4^3k_3^3} \;.
\ee

\begin{figure}[h]
\centering
\includegraphics[width=5.8in]{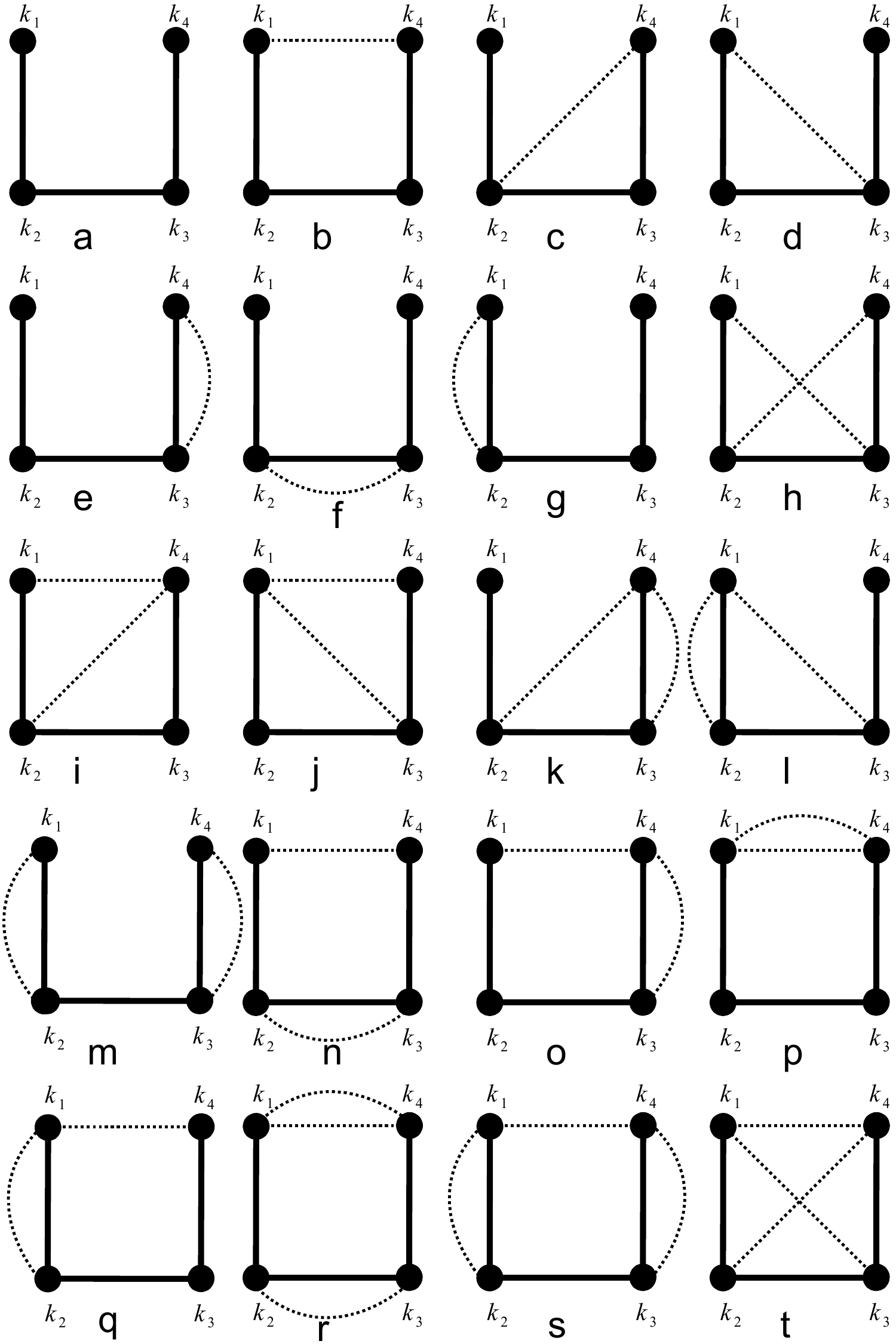}
\caption{All diagrams contributing to $\tau_{NL}$ at tree level and all corrections up to third order in $\chi$.
These diagrams contribute to $1/k_1^3 k_{12}^3 k_4^3$ term of the 4-point correlator.}
\label{TNL}
\end{figure}

\subsection{$\tau_{NL}$}
The full calculation for $\tau_{NL}$ involves integrals with poles which owing to the structure of $\tau_{NL}$ are cut off by
sums of momentum vectors. Proceeding with the same labelling conventions,
\bea
\langle \zeta_{\V{k}_1} \zeta_{\V{k}_2} \zeta_{\V{k}_3} \zeta_{\V{k}_4} \rangle_{\tau_{NL}a}
&=& A_2^2A_1^2K(2\pi^2\P)^3 \frac{1}{k_1^3k_{12}^3k_4^3}\;,
\eea
where $k_{12} \equiv |\V{k}_1+\V{k}_2|^3$. Using integral approximations already outlined (except often with 4 poles now instead of three),
\bea
\langle \zeta_{\V{k}_1} \zeta_{\V{k}_2} \zeta_{\V{k}_3} \zeta_{\V{k}_4} \rangle_{\tau_{NL}b}
&=& A_2^4K(2\pi^2)^3\P^4 \frac{\ln{({\min(k_1,k_{12},k_4)L})}}{k_1^3k_{12}^3k_4^3}
\\
\langle \zeta_{\V{k}_1} \zeta_{\V{k}_2} \zeta_{\V{k}_3} \zeta_{\V{k}_4} \rangle_{\tau_{NL}c}
&=&  A_3A_2^2A_1K(2\pi^2)^3\P^4 \frac{\ln{({\min(k_{12},k_4)L})}}{k_1^3k_{12}^3k_4^3}
\\
\langle \zeta_{\V{k}_1} \zeta_{\V{k}_2} \zeta_{\V{k}_3} \zeta_{\V{k}_4} \rangle_{\tau_{NL}d}
&=& A_3A_2^2A_1K(2\pi^2)^3\P^4 \frac{\ln{({\min(k_{12},k_1)L})}}{k_1^3k_{12}^3k_4^3}
\\
\langle \zeta_{\V{k}_1} \zeta_{\V{k}_2} \zeta_{\V{k}_3} \zeta_{\V{k}_4} \rangle_{\tau_{NL}e}
&=& A_3A_2^2A_1K(2\pi^2)^3\P^4 \frac{\ln{({k_4}L)}}{k_1^3k_{12}^3k_4^3}
\\
\langle \zeta_{\V{k}_1} \zeta_{\V{k}_2} \zeta_{\V{k}_3} \zeta_{\V{k}_4} \rangle_{\tau_{NL}f}
&=& A_3^2A_1^2K(2\pi^2)^3\P^4 \frac{\ln{({k_{12}}L)}}{k_1^3k_{12}^3k_4^3}
\\
\langle \zeta_{\V{k}_1} \zeta_{\V{k}_2} \zeta_{\V{k}_3} \zeta_{\V{k}_4} \rangle_{\tau_{NL}g}
&=& A_3A_2^2A_1K(2\pi^2)^3\P^4 \frac{\ln{({k_1}L)}}{k_1^3k_{12}^3k_4^3}
\\
\langle \zeta_{\V{k}_1} \zeta_{\V{k}_2} \zeta_{\V{k}_3} \zeta_{\V{k}_4} \rangle_{\tau_{NL}h}
&=& A_3^2A_2^2K(2\pi^2)^3\P^5 \frac{\ln{({\min(k_{12},k_4)L})}\ln{({\min(k_{12},k_1)L})}}{k_1^3k_{12}^3k_4^3}
\\
\langle \zeta_{\V{k}_1} \zeta_{\V{k}_2} \zeta_{\V{k}_3} \zeta_{\V{k}_4} \rangle_{\tau_{NL}i}
&=& A_3^2A_2^2K(2\pi^2)^3\P^5 \frac{\ln{({\min(k_{12},k_4)L})}\ln{({\min(k_1,k_{12},k_4)L})}}{k_1^3k_{12}^3k_4^3}
\nonumber\\
\\
\langle \zeta_{\V{k}_1} \zeta_{\V{k}_2} \zeta_{\V{k}_3} \zeta_{\V{k}_4} \rangle_{\tau_{NL}j}
&=& A_3^2A_2^2K(2\pi^2)^3\P^5 \frac{\ln{({\min(k_{12},k_1)L})}\ln{({\min(k_1,k_{12},k_4)L})}}{k_1^3k_{12}^3k_4^3}
\nonumber\\
\eea
\bea
\langle \zeta_{\V{k}_1} \zeta_{\V{k}_2} \zeta_{\V{k}_3} \zeta_{\V{k}_4} \rangle_{\tau_{NL}k}
&=& A_3^3A_1K(2\pi^2)^3\P^5 \frac{\ln{({k_4L})}\ln{({\min(k_{12},k_4)L})}}{k_1^3k_{12}^3k_4^3}
\\
\langle \zeta_{\V{k}_1} \zeta_{\V{k}_2} \zeta_{\V{k}_3} \zeta_{\V{k}_4} \rangle_{\tau_{NL}l}
&=& A_3^3A_1K(2\pi^2)^3\P^5 \frac{\ln{({k_1L})}\ln{({\min(k_{12},k_1)L})}}{k_1^3k_{12}^3k_4^3}
\\
\langle \zeta_{\V{k}_1} \zeta_{\V{k}_2} \zeta_{\V{k}_3} \zeta_{\V{k}_4} \rangle_{\tau_{NL}m}
&=& A_3^2A_2^2K(2\pi^2)^3\P^5 \frac{\ln{({k_1}L)}\ln{({k_4}L)}}{k_1^3k_{12}^3k_4^3}
\\
\langle \zeta_{\V{k}_1} \zeta_{\V{k}_2} \zeta_{\V{k}_3} \zeta_{\V{k}_4} \rangle_{\tau_{NL}n}
&=& A_3^2A_2^2K(2\pi^2)^3\P^5 \frac{\ln{({k_{12}L})}\ln{({\min(k_1,k_{12},k_4)L})}}{k_1^3k_{12}^3k_4^3}
\\
\langle \zeta_{\V{k}_1} \zeta_{\V{k}_2} \zeta_{\V{k}_3} \zeta_{\V{k}_4} \rangle_{\tau_{NL}o}
&=& A_3^2A_2^2K(2\pi^2)^3\P^5 \frac{\ln{({k_4L})}\ln{({\min(k_1,k_{12},k_4)L})}}{k_1^3k_{12}^3k_4^3}
\\
\langle \zeta_{\V{k}_1} \zeta_{\V{k}_2} \zeta_{\V{k}_3} \zeta_{\V{k}_4} \rangle_{\tau_{NL}p}
&=& {1 \over 2} A_3^2A_2^2K(2\pi^2)^3\P^5 \frac{\ln{({\min(k_1,k_{12},k_4)L})}^2}{k_1^3k_{12}^3k_4^3}
\\
\langle \zeta_{\V{k}_1} \zeta_{\V{k}_2} \zeta_{\V{k}_3} \zeta_{\V{k}_4} \rangle_{\tau_{NL}q}
&=& A_3^2A_2^2K(2\pi^2)^3\P^5 \frac{\ln{({k_1L})}\ln{({\min(k_1,k_{12},k_4)L})}}{k_1^3k_{12}^3k_4^3}
\\
\langle \zeta_{\V{k}_1} \zeta_{\V{k}_2} \zeta_{\V{k}_3} \zeta_{\V{k}_4} \rangle_{\tau_{NL}r}
&=& {1 \over 2}A_3^4K(2\pi^2)^3\P^6 \frac{\ln{({k_{12}L})}\ln{({\min(k_1,k_{12},k_4)L})^2}}{k_1^3k_{12}^3k_4^3}
\\
\langle \zeta_{\V{k}_1} \zeta_{\V{k}_2} \zeta_{\V{k}_3} \zeta_{\V{k}_4} \rangle_{\tau_{NL}s}
&=& A_3^4K(2\pi^2)^3\P^6 \frac{\ln{({k_{4}L})}\ln{({k_{4}L})}\ln{({\min(k_1,k_{12},k_4)L})}}{k_1^3k_{12}^3k_4^3}
\\
\langle \zeta_{\V{k}_1} \zeta_{\V{k}_2} \zeta_{\V{k}_3} \zeta_{\V{k}_4} \rangle_{\tau_{NL}t}
&=& A_3^4K(2\pi^2)^3\P^6 \frac{\ln{({\min(k_1,k_{12},k_4)L})}\ln{({\min(k_1,k_{12})L})}\ln{({\min(k_{12},k_4)L})}}{k_1^3k_{12}^3k_4^3}
\nonumber\\
\eea

%\bibliographystyle{JHEP}
%\bibliography{stringlist}
\providecommand{\href}[2]{#2}\begingroup\raggedright\endgroup

\end{document}